\documentclass{ws-ijmpa-mod}

\begin{document}

\markboth{Stefan Leupold}
{Chiral Partners in a Chirally Broken World}

%
\catchline{}{}{}{}{}
%

\title{CHIRAL PARTNERS IN A CHIRALLY BROKEN WORLD
}

\author{STEFAN LEUPOLD
}

\address{Institut f\"ur Theoretische Physik, Johann Wolfgang Goethe-Universit\"at Frankfurt, Max-von-Laue-Str.\ 1\\
D-60438 Frankfurt am Main, Germany
\\
s.leupold@gsi.de}

\author{MARKUS WAGNER}

\address{Institut f\"ur Theoretische Physik, Universit\"at Giessen, Heinrich-Buff-Ring 16\\
D-35392 Giessen, Germany}

\maketitle


\begin{abstract}
The isovector--vector and the isovector--axial-vector current are related by a chiral transformation.
These currents can be called chiral partners at the fundamental level.
In a world where chiral symmetry was not broken, the corresponding current-current
correlators would show the same spectral information. In the real world chiral symmetry is spontaneously
broken. A prominent peak --- the $\rho$-meson --- shows up in the vector spectrum 
(measured in $e^+ e^-$-collisions and $\tau$-decays). 
On the other hand, in the axial-vector spectrum a broad bump appears --- the $a_1$-meson (also accessible
in $\tau$-decays). It is tempting to call $\rho$ and $a_1$ chiral partners at the hadronic level.
Strong indications are brought forward that these ``chiral partners'' do not only differ in mass but
even in their nature: The $\rho$-meson appears dominantly as a quark-antiquark state with small
modifications from an attractive pion-pion interaction. The $a_1$-meson, on the other hand, can be 
understood as a meson-molecule state mainly formed by the attractive interaction 
between pion and $\rho$-meson. A key issue here is that the meson-meson interactions are fixed by chiral
symmetry breaking. 
It is demonstrated that one can understand the vector and the axial-vector
spectrum very well within this interpretation. It is also shown that the opposite cases, namely $\rho$ as
a pion-pion molecule or $a_1$ as a quark-antiquark state lead to less satisfying results.
Finally speculations on possible in-medium changes of hadron properties are presented.
\end{abstract}

\ccode{PACS numbers: 11.30.Rd, 12.38.-t, 14.40.Cs, 25.75.Nq}


\section[Symmetry Breaking]{Chiral Symmetry Breaking}

\begin{figure}
\centerline{
\psfig{file=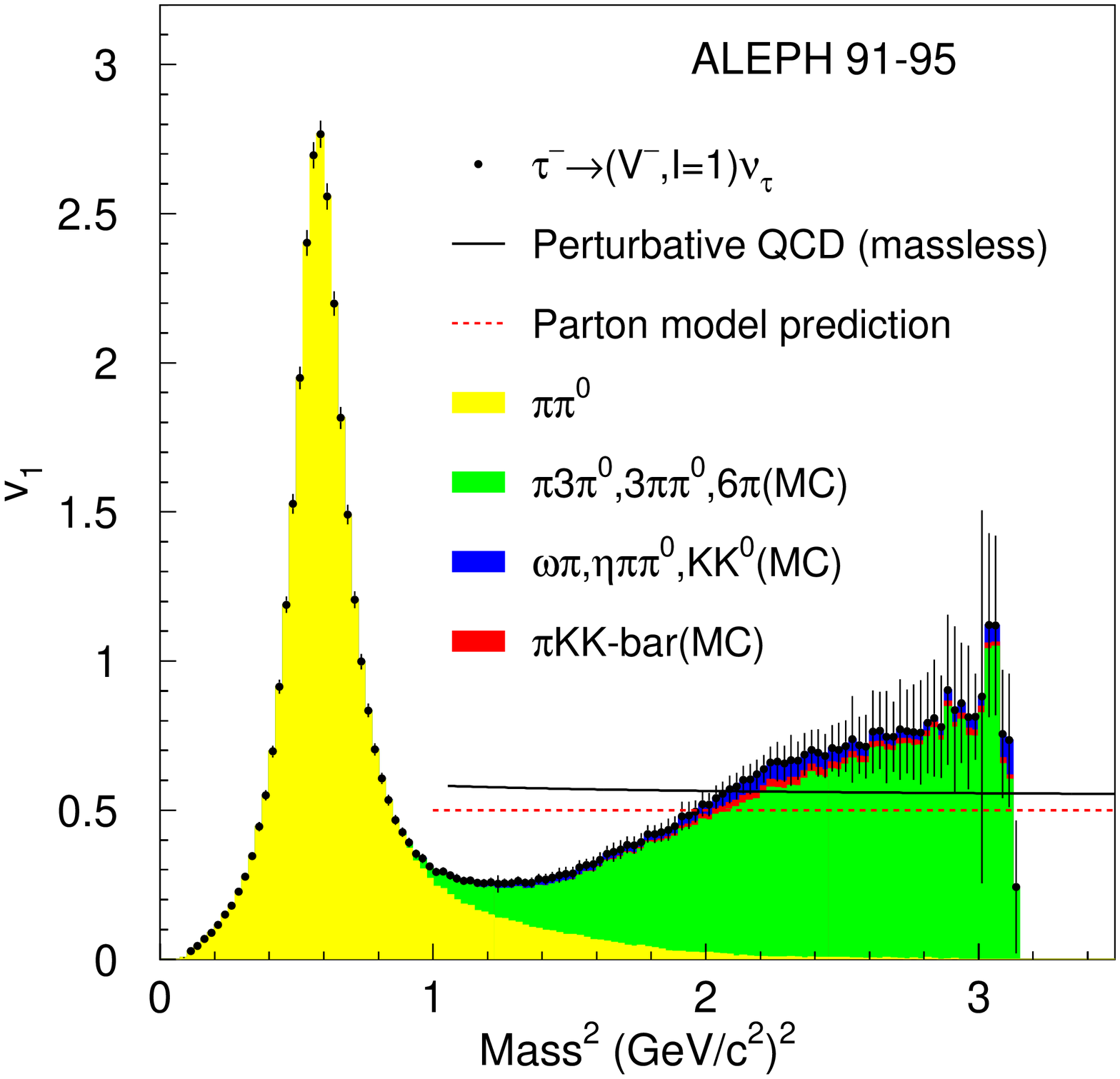,width=0.49\textwidth}
\psfig{file=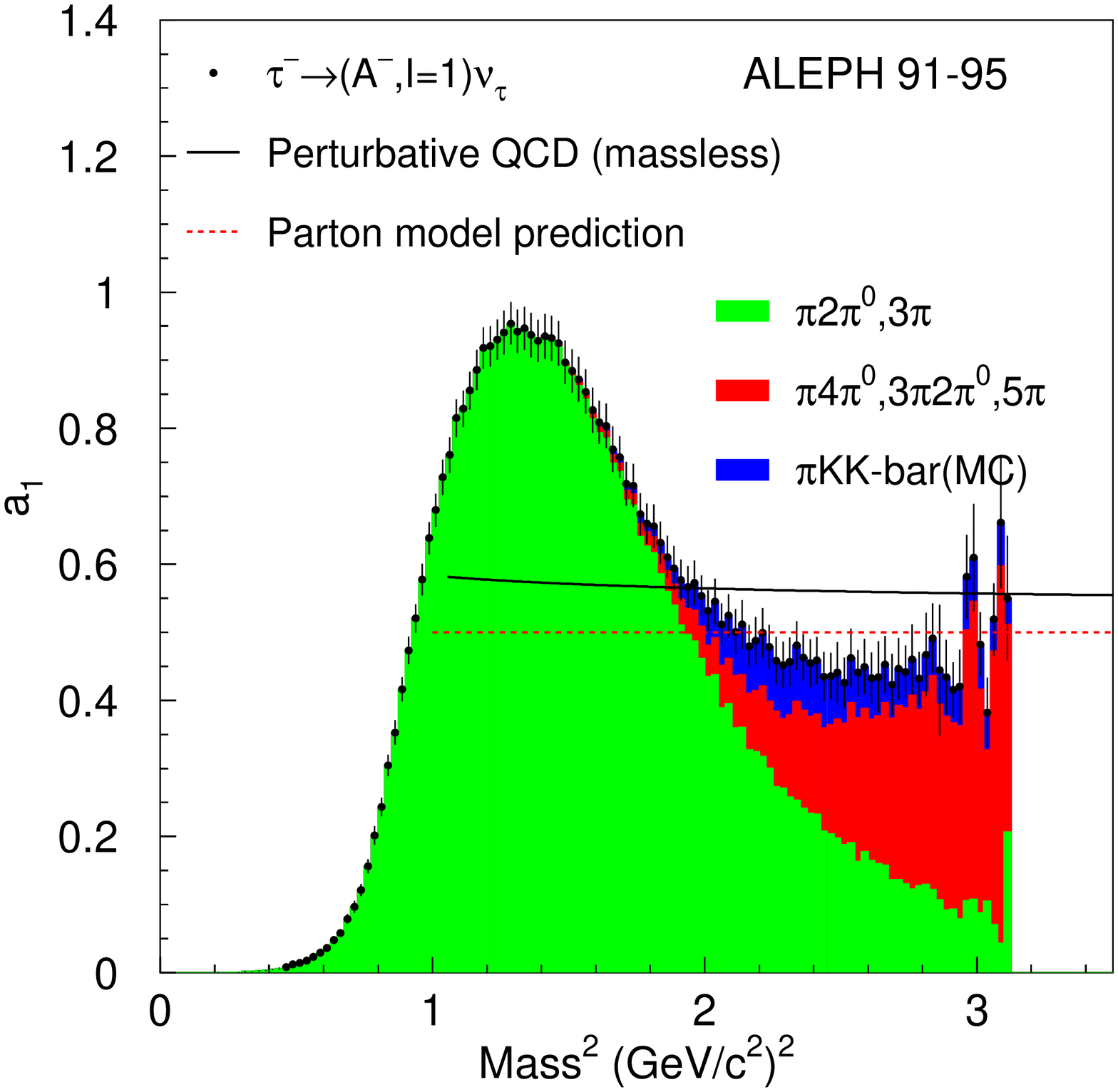,width=0.49\textwidth}
}
\caption{Spectral information of the vector ({\it left}) and axial-vector ({\it right}) 
current. Figures taken from Ref.\ \protect\refcite{aleph}. 
\label{fig:chibr-exp}}
\end{figure}

According to our present knowledge of the strong interaction, chiral symmetry is an
approximate symmetry of the interaction but not of the ground state (vacuum). 
In other words, in the
light quark sector chiral
symmetry is spontaneously broken (on top of the small explicit breaking due to the
non-vanishing quark masses).
A strong indication that chiral symmetry is spontaneously broken comes from the
study of the isovector--vector current $\vec j_V^\mu = \bar q \vec \tau \gamma^\mu q$ 
and the isovector--axial-vector current 
$\vec j_A^\mu = \bar q \vec \tau \gamma_5 \gamma^\mu q$. These currents are intertwined
by a chiral transformation (see e.g.\ Ref.\ \refcite{koch} and references therein).
It is illuminating to work out the quantum numbers of these currents:\cite{peskin}
$\vec j_V^\mu$ has the quantum numbers $I^G(J^{PC})=1^+(1^{--})$, while 
$\vec j_A^\mu$ has the quantum numbers 
$I^G(J^{PC})=1^-(1^{++})$.\footnote{Strictly speaking the quantum number of 
charge conjugation $C$ can only be assigned to the neutral currents.}
These are the quantum numbers of the $\rho$- and $a_1$-meson, respectively.\cite{pdg}
Suppose for a moment that chiral symmetry was not broken, i.e.\ realized in the same way
as e.g.\ isospin symmetry (Wigner-Weyl mode). 
In this case one would expect (nearly) identical spectra from
the current-current correlators of $\vec j_V$ and $\vec j_A$. Of course, this statement
is not limited to these two currents, but applies also to other quark currents connected
by chiral transformations. However, the particular currents $\vec j_V$ and $\vec j_A$ 
are exceptional for the following reason: They are directly accessible by electroweak 
processes. Therefore, experiment can tell us whether the ground state of the strong
interaction is in the Wigner-Weyl mode of chiral symmetry. 
For example, photons couple to the neutral currents
contained in $\vec j_V$ whereas the hadronic weak current is formed 
by $\vec j_V - \vec j_A$.
The Fourier decomposition, i.e.\ the spectral information of the current-current 
correlators of $\vec j_V$ and $\vec j_A$ is depicted in Fig.\ \ref{fig:chibr-exp}.
Obviously the spectra are not identical, not even approximately. This is one of the
clearest indications that chiral symmetry is spontaneously broken.

\section[Hadrons]{Chiral Partners at the Level of Hadrons --- $\rho$ and $a_1$?}
\label{sec:partners}

Since the currents $\vec j_V$ and $\vec j_A$ are connected to each other by chiral
transformations, it is suggestive to call them chiral partners at the fundamental level
of quark currents. But which objects should one call chiral partners at the level of 
hadrons? By inspecting Fig.\ \ref{fig:chibr-exp} we find that the $\rho$-meson
shows up as a prominent peak in the vector spectrum (left panel), while the $a_1$-meson
appears as a broad bump in the axial-vector spectrum. This suggests to call
$\rho$ and $a_1$ chiral partners. On the other hand, these states do not even have the
same mass. While it is natural that ``partner spectra'' are different 
for a system where the
symmetry is spontaneously broken, there are still states which are energetically
degenerate. For a given hadron, e.g.\ a single-particle hadron state $\vert h \rangle$, 
the two-particle state $\vert \pi(0) \, h \rangle$ is (approximately) degenerate
in energy. Here $\pi(0)$ is a soft pion state, i.e.\ with vanishing momentum. 
The Goldstone theorem demands that Goldstone bosons are massless and do not interact
for vanishing momentum. Hence the two-particle state $\vert \pi(0) \, h \rangle$ is
degenerate in energy with the single-particle state $\vert h \rangle$.\footnote{This
degeneracy is broken to some extent due to the explicit symmetry breaking by the finite
current quark masses.} Should one call $a_1$ or
a $\pi$-$\rho$ system the chiral partner of the $\rho$? One might regard
it merely as a matter of definition. On the other hand, we will demonstrate 
in Sec.\ \ref{sec:nature} that these two possibilities are actually 
intimately connected. In addition, the question of chiral partners becomes a real issue
in a strongly interacting medium where chiral symmetry restoration is expected to take 
place. There, the spectra which are shown for vacuum in Fig.\ \ref{fig:chibr-exp} must 
become degenerate. If there were still peak or bump structures visible in the spectra
at the point of chiral restoration, then it would be suggestive to trace back from
which vacuum structures these in-medium structures emerge. While the in-medium structures
must show a degeneracy (chiral restoration!), the corresponding vacuum structures
do not. But it is then tempting to call the vacuum structures chiral partners as they
become degenerate at the point of chiral restoration.
We will come back to this point in Sec.\ \ref{sec:restor} below.

\section[Nature of Resonances]{Different Nature of Hadronic Chiral Partners}
\label{sec:nature}

\begin{figure}
\centerline{\psfig{file=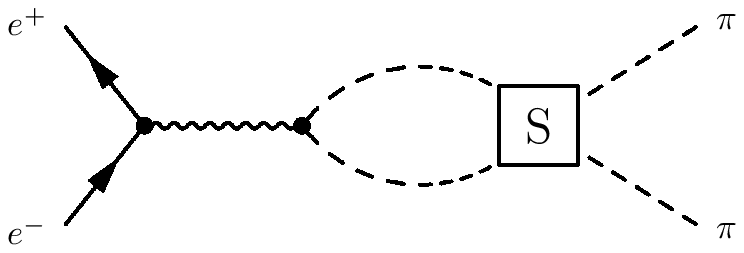,width=0.45\textwidth}   \hfill
\psfig{file=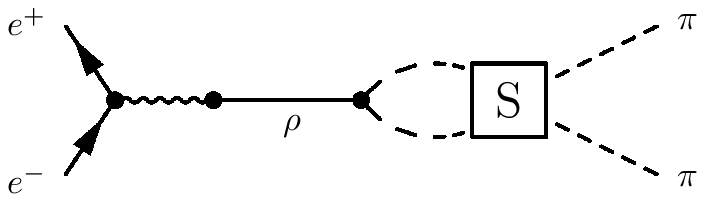,width=0.45\textwidth}
} 
\vspace*{1em}
\centerline{\psfig{file=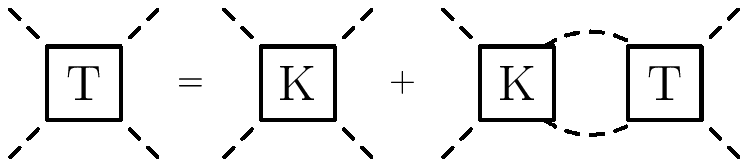,width=0.4\textwidth}} 
\vspace*{1em}
\centerline{\psfig{file=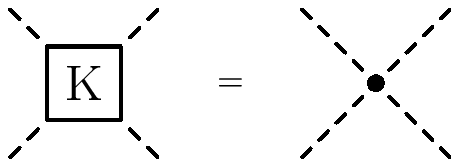,width=0.22\textwidth}  \hfill
\psfig{file=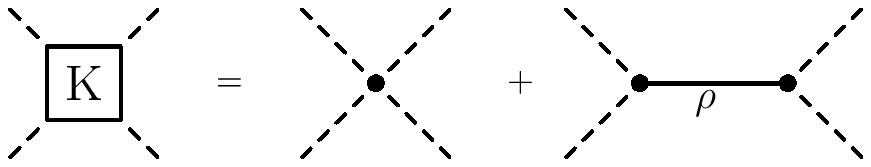,width=0.45\textwidth}
}
\vspace*{4pt}
\caption{Description of the electromagnetic form factor of the pion within the two 
scenarios. The processes from which the form factor is extracted are depicted in the top panels. 
For the first scenario (only final-state rescattering) only
the top left diagram enters. For the second scenario both diagrams in the top line are considered.
The boxes labeled with $S$ denote the $S$-matrix of pion-pion scattering. It is obtained from the 
$T$-matrix which in turn results from the
solution of a Bethe-Salpeter equation (middle panel). The kernel $K$ of the Bethe-Salpeter equation
for the first/second scenario is shown in the bottom left/right panel. In the first scenario this kernel
is fixed by the lowest order chiral interaction. It is a point interaction as 
depicted in the lower panels. In addition, for the second scenario the preformed resonance 
appears in the kernel. 
\label{fig:eepipiS}}
\end{figure}

\begin{figure}
\centerline{
\psfig{file=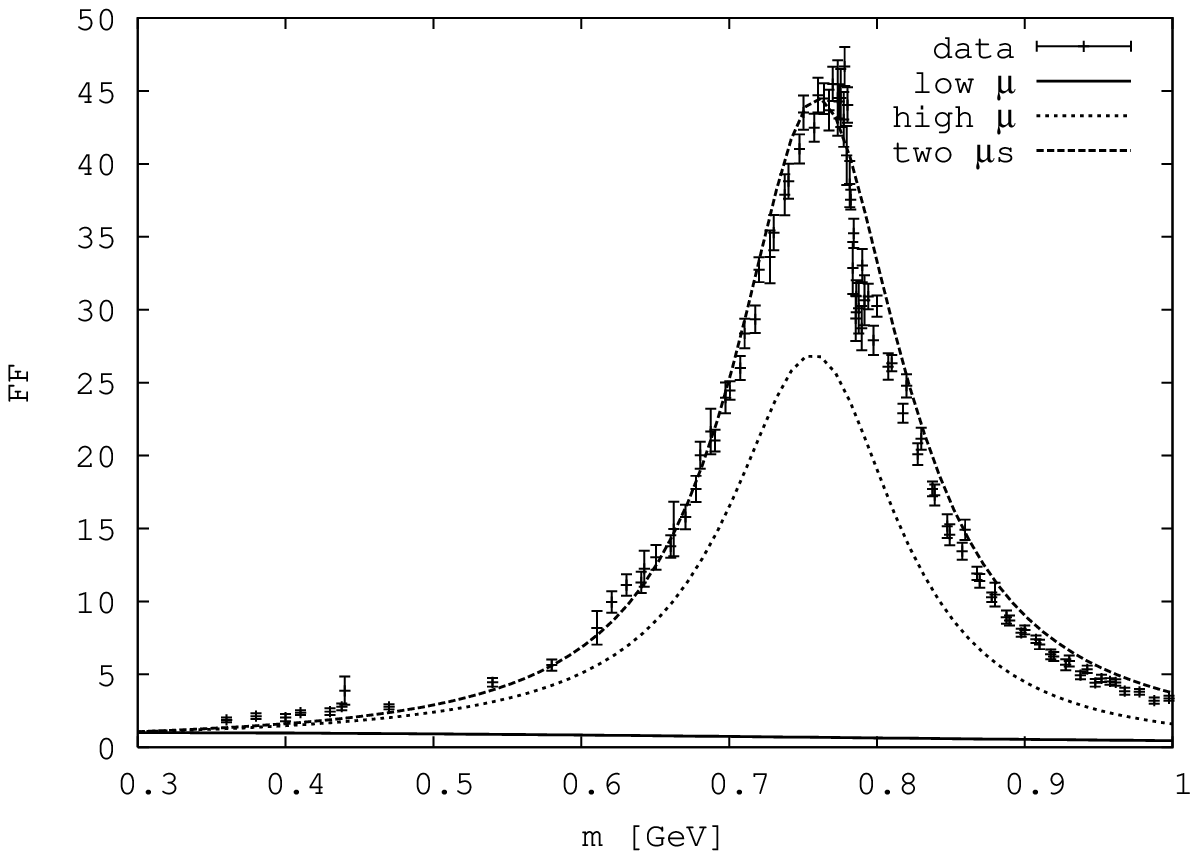,width=0.51\textwidth}
\psfig{file=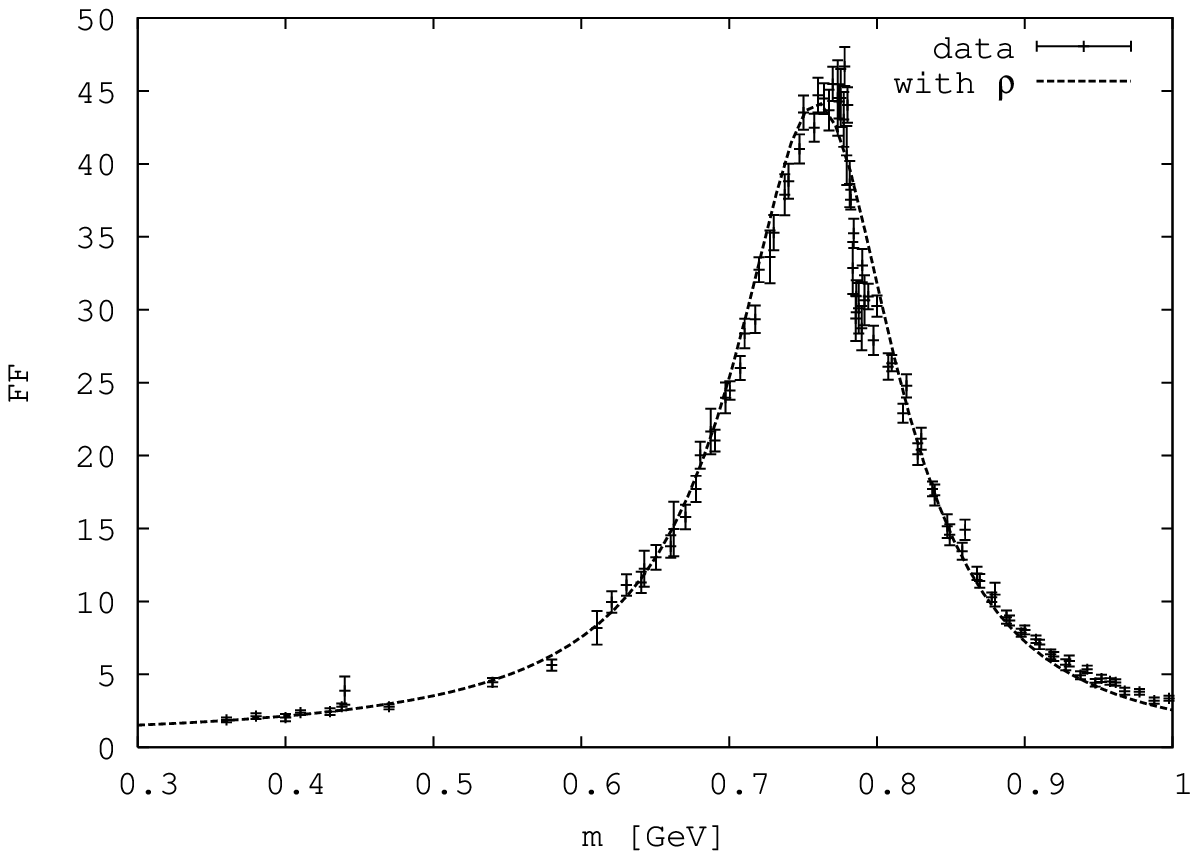,width=0.51\textwidth}
}
\caption{{\it Left:} The pion form factor in the first scenario (only rescattering of pions). The 
physically reasonable 
calculation is shown by the full line. The other calculations are technically possible, but physically
unreasonable, since they correspond to renormalization points in the TeV range. See main text for details.
{\it Right:} The pion form factor in the second scenario where an elementary 
resonance is included in addition.{\protect\cite{leupold08}} 
Data taken from Ref.\ \protect\refcite{pionformfac}.
\label{fig:norho}}
\end{figure}

\begin{figure}
\centerline{
\psfig{file=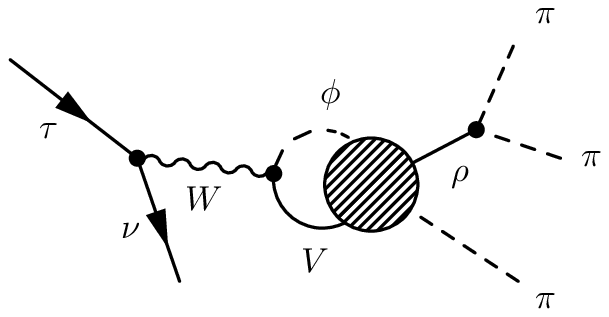,width=0.4\textwidth} \hfill
\psfig{file=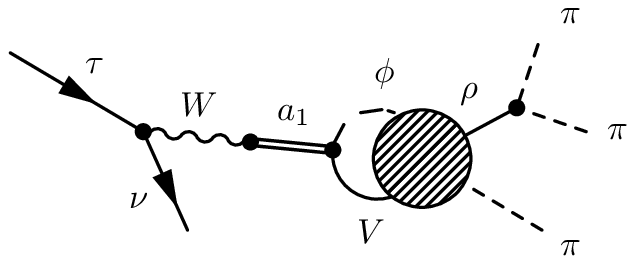,width=0.45\textwidth}
}
\vspace*{1em}
\centerline{\psfig{file=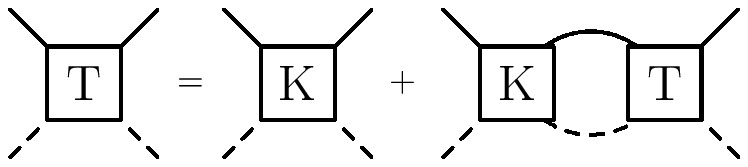,width=0.4\textwidth}} 
\vspace*{1em}
\centerline{
\psfig{file=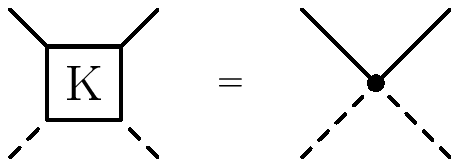,width=0.22\textwidth} \hfill
\psfig{file=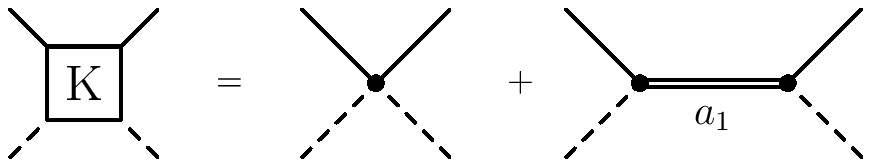,width=0.45\textwidth}
}
\caption{Diagrams relevant for the description of the process $\tau \to \nu_\tau + 3 \pi$.
For details see Fig.\ \protect\ref{fig:eepipiS}, main text and 
Ref.\ \protect\refcite{wagner-long}. \label{fig:taudecpic7}}
\end{figure}

\begin{figure}
\centerline{\psfig{file=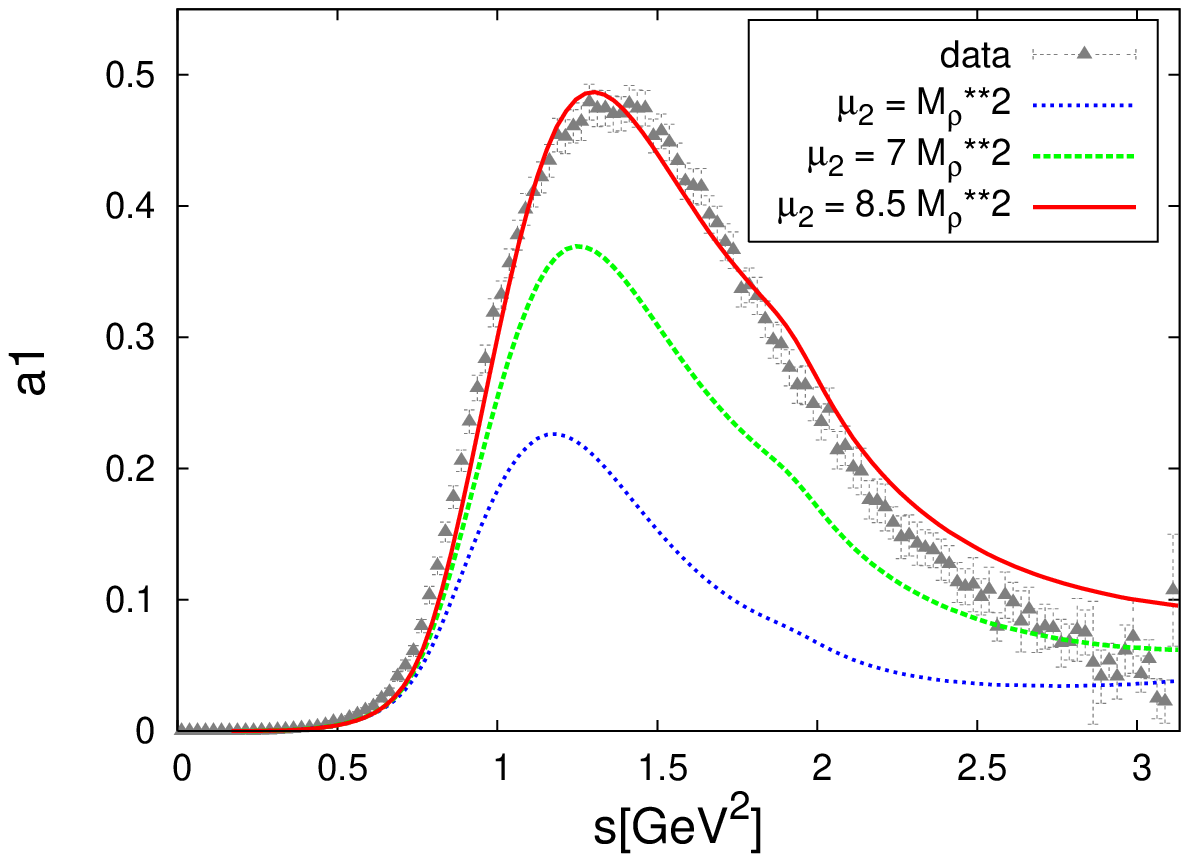,width=0.51\textwidth}
\psfig{file=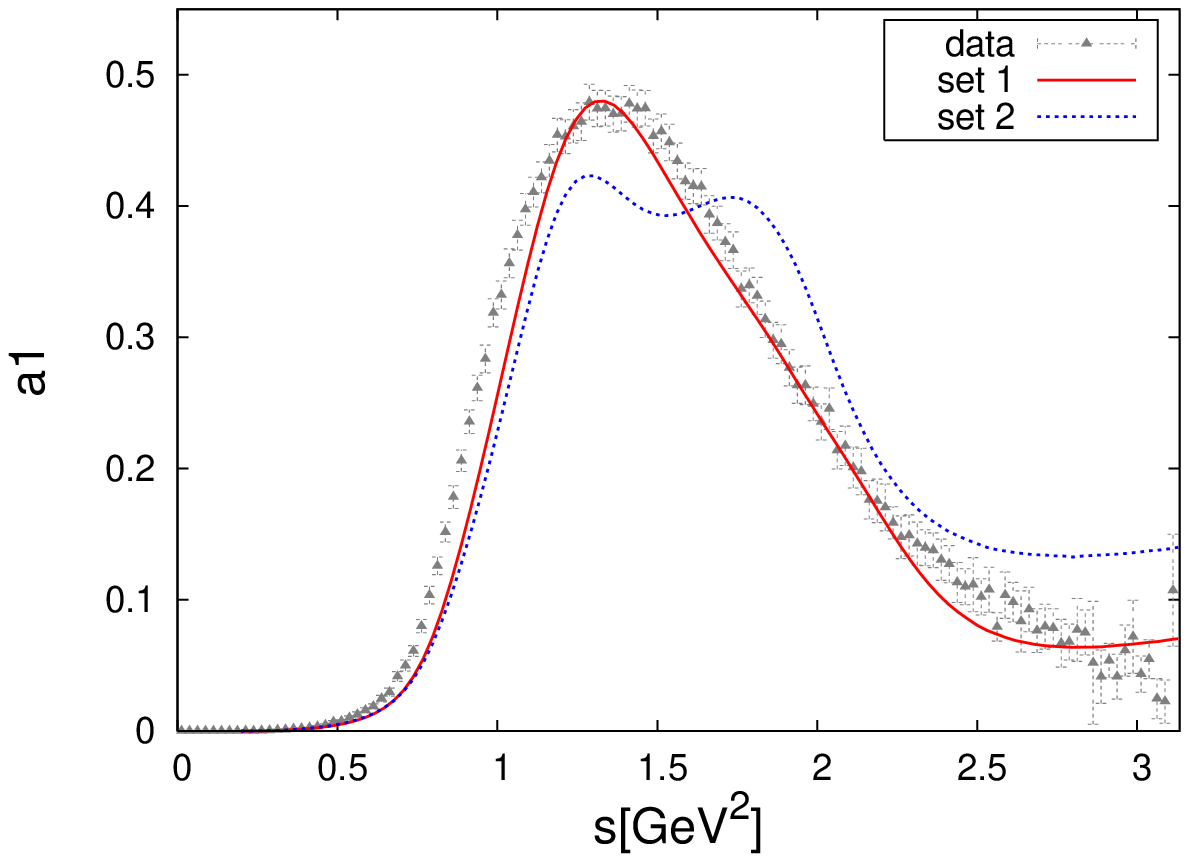,width=0.51\textwidth}
}
\caption{The axial-vector spectral information in the three-pion final state described by the first
({\it left}) and the second ({\it right}) scenario. 
See main text and Ref.\ \protect\refcite{wagner-long} for details.
Data taken from Ref.\ \protect\refcite{aleph}.
\label{fig:markus-tau-WT}}
\end{figure}

We are aiming at an understanding of the respective low-energy part of the spectra
depicted in Fig.\ \ref{fig:chibr-exp}. Both show a resonant structure: In the vector
spectrum (left panel) there is a peak at about 770 MeV called the $\rho$-meson. 
In the axial-vector spectrum (right panel) 
there is a broad bump at about 1250 MeV called the 
$a_1$-meson. Actually both low-energy parts are governed by a (quasi-)two-particle 
final state --- $\pi\pi$ for the vector and $\rho\pi$ for the 
axial-vector channel. The latter can be deduced from a Dalitz plot
analysis of the three-pion final state.\cite{wagner-long}
The general strategy is the following: 
The two-particle state is subject to final-state 
interactions (rescattering). We want to figure out whether this final-state interaction
is sufficient to create the respective resonant structure seen in 
Fig.\ \ref{fig:chibr-exp} or whether one needs in addition a preformed resonance, 
i.e.\ an elementary hadronic state
which microscopically should be regarded as a quark-antiquark state. This intrinsic 
structure is, however, not resolved at the hadronic level. We study two scenarios:
For the first scenario we only include the final-state interactions which we describe
via a Bethe-Salpeter equation. The kernel is taken from the lowest order 
chiral interaction. It is important to note that the strength of this
final-state interaction is fixed by chiral symmetry breaking and is therefore 
parameter free. In the second scenario we include in addition a preformed resonance.
If we got a reasonable description of the data from the first scenario, 
we would conclude that the resonance in the considered channel is a dynamically generated 
state, a meson-meson molecule.
Otherwise, we would conclude that the resonance has a non-negligible or even dominant
quark-antiquark contribution which can be quantified in the second scenario.
\\
{\bf Nature of resonances --- the $\rho$-meson:} 
Instead of the two-pion spectrum of Fig.\ \ref{fig:chibr-exp}, left, we study
the electromagnetic form factor of the pion in the time like region.\cite{pionformfac} 
Due to isospin symmetry this contains the same information.\cite{aleph} 
The relevant processes are depicted in Fig.\ \ref{fig:eepipiS}. 
The lowest-order chiral interaction of the two-pion system is given by the 
non-linear sigma model.\cite{gasser-leut} 
We have two parameters at our disposal: the renormalization points (a) for the loop of the transition
from the virtual photon to pions (Fig.\ \ref{fig:eepipiS}, top left) and 
(b) for the loop appearing in
the Bethe-Salpeter equation (Fig.\ \ref{fig:eepipiS}, middle).
These renormalization points are not (completely) free, however: First of all, both have to be in a 
reasonable range (see below). Second, the renormalization point for the loop in the Bethe-Salpeter equation
can be fixed by the requirement of approximate crossing symmetry\cite{lutz-kolo} 
(see also Ref.\ \refcite{hyodo} for a different line of reasoning which yields the same result).
The result of the first scenario (only final-state interaction) is shown in 
Fig.\ \ref{fig:norho}, left.
The full line labeled ``low $\mu$'' is obtained for both renormalization points set to the 
pion mass. Obviously, one fails to describe the data. 
If the renormalization points are increased, it is possible to create a peak structure.
The dotted line labeled ``high $\mu$'' is obtained for both renormalization points set to 1.1 TeV.
Finally, one gets the dashed line labeled ``two $\mu$s'' by setting the first renormalization point
(photon-to-pion transition) to 10 TeV and the second one to 1.1 TeV. Thus, from a purely technical 
point of view the approach allows for a description of the
data (dashed line). From the physical point of view, however, it must be stressed that only the full line
corresponds to a reasonable calculation, since the renormalization points should lie in a reasonable
range and not orders of magnitude away from typical hadronic scales. We conclude that with a physically
reasonable choice of parameters one cannot explain the pion form factor within a scenario which includes
only pion-pion rescattering. One needs in addition an elementary resonance as we will show next.
A similar conclusion has been drawn in Ref.\ \refcite{oller} studying the
pion-pion scattering phase shifts.
We now turn to the second scenario where an elementary resonance is included in addition to the pure 
rescattering studied in the first scenario. The form factor is now obtained from the sum of diagrams
shown in the top line of Fig.\ \ref{fig:eepipiS}. The Bethe-Salpeter equation
is formally unchanged, Fig.\ \ref{fig:eepipiS}, middle, but the kernel is now given by the sum
of the point interaction obtained from the non-linear sigma model and the elementary resonance, 
cf.\ Fig.\ \ref{fig:eepipiS}, bottom left.
As additional parameters one has now the mass of the elementary
resonance and its couplings to the photon and to two pions. Actually, changes in the renormalization points
can be compensated by changes in these resonance parameters. Therefore, one has effectively three free
parameters. As shown in Fig.\ \ref{fig:norho}, right, one gets an excellent description of the pion
form factor.\cite{leupold08} In particular, there is no two-peak structure in the theory curve since the
pion contact interaction alone is not very strong, as already shown by the full line in 
Fig.\ \ref{fig:norho}, left.
We conclude that the $\rho$-meson is dominantly a preformed (i.e.\ quark-antiquark) state with a small
two-pion admixture and not a pion-pion molecule.
\\
{\bf Nature of resonances --- the $a_1$-meson:}
The analysis of the $a_1$-meson exactly resembles the one presented for the $\rho$-meson, but the result
will be just the opposite: We will show in the following that the $a_1$-meson can be understood as a
$\pi$-$\rho$ molecule. 
The relevant processes for the description of the decay $\tau \to \nu_\tau + 3 \pi$ are schematically 
depicted in Fig.\ \ref{fig:taudecpic7}. 
The lowest-order chiral interaction of the $\rho$-$\pi$ system is given by the 
Weinberg-Tomozawa interaction.\cite{WT}
Results of the calculations for both scenarios are compared to data 
in Fig.\ \ref{fig:markus-tau-WT}. 
In the first scenario the $a_1$-meson emerges as a dynamically
generated state from final-state interactions of vector and pseudoscalar mesons. That axial-vector mesons
can be created in this way has been suggested in Ref.\ \refcite{lutz-kolo} and 
later in Ref.\ \refcite{oset-a1}.
In these works a coupled-channel treatment of $\pi$-$\rho$ and $K$-$K^*$ has been presented. We follow
this approach, but note in passing that the strangeness channel is not very important for the 
$a_1$-meson.\cite{wagner-long} For the first scenario we take the parameter-free scattering amplitude from 
Ref.\ \refcite{lutz-kolo}. Then we are left with only one free parameter, the renormalization point $\mu_2$
of the entrance loop from the $W$-boson to hadrons (cf. Fig.\ \ref{fig:taudecpic7}, top left).
We recall that this parameter should be in a reasonable range. By only tuning $\mu_2$ we get a decent
description of the data as shown in Fig.\ \ref{fig:markus-tau-WT}, left. This means that we can 
essentially describe the position, height and width of the $a_1$-bump with one parameter 
which is in the 
GeV range (and not in the TeV range as for the case of the $\rho$-meson). 
In the second scenario where we include in addition 
an elementary resonance we typically generate a double peak structure (dotted line in 
Fig.\ \ref{fig:markus-tau-WT}, right). This is not surprising since we know from the first scenario
that the final-state interaction
between $\rho$ and $\pi$ is strong enough to create a resonance dynamically. An additional elementary
resonance can only be hidden, if its coupling to the $\rho$-$\pi$ system is weak 
(which essentially brings back the first scenario) or if its mass is fine-tuned such that it appears
at the position of the dynamically generated resonance. The latter possibility is shown as the full line
in Fig.\ \ref{fig:markus-tau-WT}, right. While this is technically possible we regard it as rather
unnatural that a quark-antiquark and a meson-meson state appear at the very same position. Therefore, the
natural explanation of the $\tau$-decay data shown in Fig.\ \ref{fig:markus-tau-WT}
is that the $a_1$-meson is a dynamically
generated state, i.e.\ a meson-meson molecule\cite{wagner-long} as suggested in 
Refs.\ \refcite{lutz-kolo,oset-a1}.
To summarize the present section: There are strong indications that the $\rho$-meson and 
the $a_1$-meson, 
the
``chiral partners'' at the level of hadrons, are not only different in mass, but actually different in 
nature: The $\rho$-meson is dominantly a quark-antiquark state 
whereas the $a_1$-meson is dominantly a
meson-meson state (mostly $\rho$-$\pi$).

\section[Chiral Restoration]{Outlook to Chiral Restoration}
\label{sec:restor}

As already discussed in Sec.\ \ref{sec:partners} we expect that the spectral information of the 
vector and the axial-vector
current become identical at the point of chiral restoration. There are various scenarios conceivable
how this degenerate in-medium spectrum might look like. Here we briefly discuss only two.
The degeneracy scenario: 
We have seen above that the $\rho$-meson is dominantly a single-particle state at the hadronic level
(and not a pion-pion correlation). If the $\rho$-meson was still dominantly a single-particle state
at the point of chiral restoration (i.e.\ if it still showed up as a prominent peak in the spectrum), 
this would require the existence of another single-particle state 
at the hadronic level with opposite parity, i.e.\ an axial-vector state. Since we have shown that
the $a_1$-meson is well described as a two-particle state, a $\rho$-$\pi$ correlation, there should be
another, i.e.\ higher-lying axial-vector state which becomes degenerate with the $\rho$-meson.
We cannot exclude this possibility and within our formalism we have not much to say about this scenario. However, there is at least one alternative.
\begin{figure}
\centerline{\psfig{file=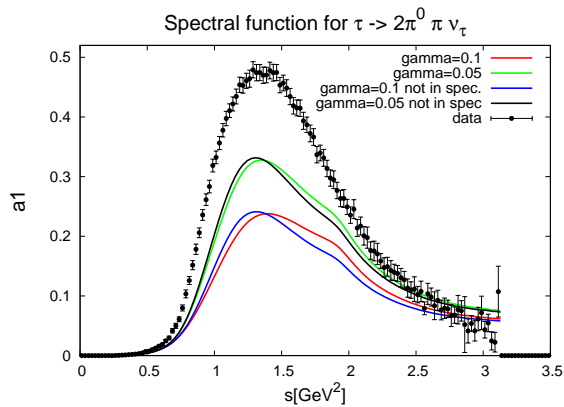,width=0.6\textwidth}}
\caption{Simple model for the in-medium axial-vector spectral function. The $\rho$-meson
propagator is modified by an additional constant in-medium width of 50 MeV (upper curves)
or 100 MeV (lower curves), respectively. (The difference between the respective curves close by
to each other is irrelevant for the present purpose.) 
Data taken from Ref.\ \protect\refcite{aleph}.
\label{fig:in-med-a1}}
\end{figure}
The melting scenario: It might appear that the $\rho$-meson dissolves already in 
hadronic matter which can be interpreted as a precursor to deconfinement. 
Then also the $a_1$-meson should dissolve.
In principle, this can be tested in our approach. In the following we present a very simple model:
We increase the width of the $\rho$-meson by a constant (by 50 or 100 MeV, respectively) and study
what happens to the dynamically generated $a_1$. It must be stressed that this model should be regarded
as a precursor to more serious calculations. E.g., 
a realistic in-medium width of the $\rho$-meson would
not be independent of the momentum of the $\rho$-meson relative to the medium.\cite{post}
In addition, one also expects a strong in-medium effect on the pion and not only on the $\rho$-meson.\cite{post}
These aspects are not covered by the simple model studied here. The result is shown in 
Fig.\ \ref{fig:in-med-a1}. Obviously, the $a_1$-meson also melts, if the $\rho$-meson melts.
This does not prove that the melting scenario is the correct approach to chiral restoration, but
at least we obtain a consistent picture. In a somewhat sloppy way one might say, 
that the problem
of the missing partner of the $\rho$-meson on the single-particle level is solved by deconfinement.


\end{document}